\documentclass[preprint,prc,showpacs,showkeys]{revtex4}
\usepackage{amsfonts}
\usepackage{amsmath}
\usepackage{amssymb}
\usepackage{graphicx}
\usepackage{rotating}

\providecommand{\U}[1]{\protect\rule{.1in}{.1in}}
\providecommand{\U}[1]{\protect\rule{.1in}{.1in}}
\providecommand{\U}[1]{\protect\rule{.1in}{.1in}}

\begin{document}

\title{Electron assisted neutron exchange process in solid state environment%
\\
}
\author{P\'{e}ter K\'{a}lm\'{a}n\footnote{%
retired from Budapest University of Technology and Economics, Institute of
Physics, \newline
e-mail: kalmanpeter3@gmail.com}}
\author{Tam\'{a}s Keszthelyi}
\affiliation{Budapest University of Technology and Economics, Institute of Physics,
Budafoki \'{u}t 8. F., H-1521 Budapest, Hungary\ }

\begin{abstract}
Electron assisted neutron exchange process in solid state environment is
investigated. It is shown that if a metal is irradiated with free electrons
then the $e+$ $_{Z}^{A_{1}}X+$ $_{Z}^{A_{2}}X\rightarrow e^{\prime }+$ $%
_{Z}^{A_{1}-1}X+$ $_{Z}^{A_{2}+1}X+\Delta $ electron assisted neutron
exchange process has measurable probability even in the case of slow
electrons of energy much less than the reaction energy $\Delta $. The
transition probability per unit time, the cross section of the process and
the yield in an irradiated sample are determined in the Weisskopf and long
wavelength approximations and in the single particle shell model. Numerical
data for the $e+$ $_{28}^{A_{1}}Ni+$ $_{28}^{A_{2}}Ni\rightarrow e^{\prime
}+ $ $_{28}^{A_{1}+1}Ni+$ $_{28}^{A_{2}-1}Ni+\Delta $ and the $e+$ $%
_{46}^{A_{1}}Pd+$ $_{46}^{A_{2}}Pd\rightarrow e^{\prime }+$ $%
_{46}^{A_{1}+1}Pd+$ $_{46}^{A_{2}-1}Pd+\Delta $ electron assisted neutron
exchange reactions are also presented.
\end{abstract}

\volumenumber{number}
\issuenumber{number}
\eid{identifier}
\startpage{1}
\endpage{}
\maketitle

\section{Introduction}

In the last two decades extraordinary observations were made in low energy
accelerator physics. Namely the so called anomalous screening effect was
observed investigating astrophysical factors of nuclear reactions of low
atomic numbers mostly in deuterated metal targets \cite{Raiola1}.

Motivated by these observations we have searched physical processes that may
effect nuclear reactions in solid state environment. We theoretically found 
\cite{kk2} that the leading channel of the $p+d\rightarrow $ $^{3}He$
reaction in solid environment is the so called solid state internal
conversion process, an adapted version of ordinary internal conversion
process \cite{Hamilton}. It was shown \cite{kk2} that if the reaction $%
p+d\rightarrow $ $^{3}He$ takes place in solid material the nuclear energy
is taken away by an electron of the environment instead of the emission of a 
$\gamma $ photon.

These observations raise the question of the possibility of further
modification of nuclear processes due to solid state environment. In this
paper the electron assisted neutron exchange process is discussed in solid
state environment.

Let us consider the following general nuclear reaction

\begin{equation}
\text{ }_{Z_{1}}^{A_{1}}X+\text{ }_{Z_{2}}^{A_{2}}Y\rightarrow \text{ }%
_{Z_{1}}^{A_{1}-1}X+\text{ }_{Z_{2}}^{A_{2}+1}Y+\Delta ,  \label{exchange}
\end{equation}%
which is called neutron exchange reaction further on. Here $\Delta $ is the
energy of the reaction, i.e. the difference between the rest energies of the
initial $\left( _{Z_{1}}^{A_{1}}X+_{Z_{2}}^{A_{2}}Y\right) $ and final $%
\left( _{Z_{1}}^{A_{1}-1}X+\text{ }_{Z_{2}}^{A_{2}+1}Y\right) $ states. In $%
\left( \ref{exchange}\right) $ the $_{Z_{1}}^{A_{1}}X$ nucleus loses a
neutron which is taken up by the $_{Z_{2}}^{A_{2}}Y$ nucleus. The process is
energetically forbidden if $\Delta <0$. If the relative energy of the two
initial nuclei is high enough to bring them within or near to the range of
the nuclear force process $\left( \ref{exchange}\right) $ usually takes
place spontaneously. It is also a process of type $\left( \ref{exchange}%
\right) $ if $Z_{1}=Z_{2}=Z$, i.e. the process%
\begin{equation}
_{Z}^{A_{1}}X+\text{ }_{Z}^{A_{2}}X\rightarrow \text{ }_{Z}^{A_{1}-1}X+\text{
}_{Z}^{A_{2}+1}X+\Delta  \label{exchange 0}
\end{equation}%
which is considered further on. One possible realization of process $\left( %
\ref{exchange 0}\right) $ is if the beam consists of particles $%
_{Z}^{A_{1}}X $ and particles $_{Z}^{A_{2}}X$ are targets.

However, if the energy of the beam is less than the neutron separation
energy then the Coulomb interaction between projectile and target nuclei
creates a virtual neutron which is captured by the other nucleus. So in this
case the process can be considered as a second order process from the point
of view of perturbation calculation.

The cross section of process $\left( \ref{exchange 0}\right) $ can be\
derived applying the Coulomb solution $\varphi (\mathbf{r})$, which is the
wave function of a free particle of charge number $Z$ in a repulsive Coulomb
field of charge number $Z$ \cite{Alder}, in the description of relative
motion of projectile and target. Since $\varphi (\mathbf{r})\sim e^{-\pi
\eta /2}\Gamma (1+i\eta )$, the cross section of the process is proportional
to 
\begin{equation}
\left\vert e^{-\pi \eta /2}\Gamma (1+i\eta )\right\vert ^{2}=\frac{2\pi \eta
\left( E\right) }{\exp \left[ 2\pi \eta \left( E\right) \right] -1}=F_{C}(E).
\label{Fjk}
\end{equation}%
Here $\Gamma $ is the Gamma function and $\eta $ is the Sommerfeld parameter
which reads as 
\begin{equation}
\eta \left( E\right) =Z^{2}\alpha _{f}\sqrt{\left( \frac{A_{1}A_{2}}{%
A_{1}+A_{2}}\right) \frac{m_{0}c^{2}}{2E}}  \label{eta23}
\end{equation}%
in the case of colliding particles of charge numbers $Z_{1}=Z_{2}=Z$ and
rest masses $m_{1}=A_{1}m_{0}$ and $m_{2}=A_{2}m_{0}$. $E$ is the kinetic
energy in the center of mass coordinate system, $m_{0}c^{2}=931.494$ $MeV$
is the atomic energy unit and $\alpha _{f}$ is the fine structure constant.
Thus it is a fact that the rate of the nuclear reaction $\left( \ref%
{exchange 0}\right) $ becomes very small at low energies as a consequence of 
$F_{C}(E)$ being small.

The electron assisted version of process $\left( \ref{exchange 0}\right) $
is 
\begin{equation}
e+\text{ }_{Z}^{A_{1}}X+\text{ }_{Z}^{A_{2}}X\rightarrow e^{\prime }+\text{ }%
_{Z}^{A_{1}-1}X+\text{ }_{Z}^{A_{2}+1}X+\Delta  \label{exchange 10}
\end{equation}%
where $e$ and $e^{\prime }$ denote electron. It is also a second order
process in which the electron Coulomb interacts with the $_{Z}^{A_{1}}X$
nucleus, the intermediate, virtual neutron and the $_{Z}^{A_{1}-1}X$ nucleus
are created due to this interaction and the intermediate, virtual neutron is
captured due to the strong interaction by the nucleus $_{Z}^{A_{2}}X$
forming the nucleus $_{Z}^{A_{2}+1}X$ in this manner.

The physical background of the virtual neutron stripping due to the Coulomb
interaction is worth mentioning. The attractive Coulomb interaction acts
between the $Z$ protons and the electron. The neutrons do not feel Coulomb
interaction. So one can say that in fact the nucleus $_{Z}^{A_{1}-1}X$ is
stripped of the neutron due to the Coulomb attraction.

When describing the effect of the Coulomb interaction between the nucleus of
charge number $Z$ and a slow electron one can also use Coulomb function and,
consequently, the cross section of process $\left( \ref{exchange 10}\right) $
to be investigated is proportional to 
\begin{equation}
F_{e}(E)=\frac{2\pi \eta _{e}\left( E\right) }{\exp \left[ 2\pi \eta
_{e}\left( E\right) \right] -1}  \label{Fe}
\end{equation}%
but with 
\begin{equation}
\eta _{e}=-Z\alpha _{f}\sqrt{\frac{m_{e}c^{2}}{2E}}.  \label{etae}
\end{equation}%
Here $m_{e}$ is the rest mass of the electron. In the case of low (less than 
$0.1$ $MeV$) kinetic energy of the electron $F_{e}(E)$ reads approximately
as $F_{e}(E)=\left\vert 2\pi \eta _{e}\left( E\right) \right\vert $. If we
compare the cross sections of processes $\left( \ref{exchange 10}\right) $
and $\left( \ref{exchange 0}\right) $ their ratio is proportional to $%
F_{e}(E)/F_{C}(E)\simeq 1/F_{C}(E)\gg 1$. Therefore for small $E$ process $%
\left( \ref{exchange 10}\right) $ is preferred to process $\left( \ref%
{exchange 0}\right) $ since $1/F_{C}(E)$ becomes extremely large with
decreasing $E$. The cross section of the electron assisted neutron exchange
process has a further increase due to the large number density $\simeq
1/d^{3}$ of the two types of nuclei in the solid to which the cross section
is also proportional as it will be seen. Here $d$ is the lattice parameter
of order of magnitude of $10^{-8}$ $cm$.

We investigate systems in which process $\left( \ref{exchange 10}\right) $
can take place. The solid is a metal (e.g. $Ni$ or $Pd$) which contains the
nuclei $_{Z}^{A_{1}}X$ and $_{Z}^{A_{2}}X$ and it is irradiated with slow,
free electrons (of nonrelativistic energy). The nuclei $_{Z}^{A_{1}}X$ (or $%
_{Z}^{A_{2}}X$) interact with the ingoing free electrons via Coulomb
interaction.

\section{Electron assisted neutron exchange process}

Let us take a solid (in our case a metal) which is irradiated by a
monoenergetic beam of slow, free electrons. The corresponding sub-system
Hamiltonians are $H_{solid}$ and $H_{e}$. It is supposed that their
eigenvalue problems are solved, and the complete set of the eigenvectors of
the two independent systems are known. The interaction between them is the
Coulomb interaction of potential $V^{Cb}\left( \mathbf{x}\right) $ and the
other interaction that is taken into account between the nucleons of the
solid is the strong interaction potential $V^{St}\left( \mathbf{x}\right) $.
In the second order process investigated an electron takes part in a Coulomb
scattering with an atomic nucleus of the solid. In the intermediate state a
\ virtual free neutron $n$ is created which is captured due to the strong
interaction with some other nucleus of the solid. The reaction energy $%
\Delta $ is shared between the quasi-free final electron and the two final
nuclei which take part in the process. Since the aim of this paper is to
show the fundamentals of the main effect, the simplest description is chosen.

The electron of charge $-e$ and the nucleus $_{Z}^{A_{1}}X$ of charge $Ze$
take part in Coulomb-interaction. We use a screened Coulomb potential of the
form%
\begin{equation}
V^{Cb}\left( \mathbf{x}\right) =-\frac{e^{2}Z}{2\pi ^{2}}\int \frac{1}{%
q^{2}+\lambda ^{2}}\exp \left( i\mathbf{q}\cdot \mathbf{x}\right) d\mathbf{q}
\label{Vcb1}
\end{equation}%
with screening parameter $\lambda $ and coupling strength $e^{2}=\alpha
_{f}\hbar c$. For the strong interaction the interaction potential 
\begin{equation}
V^{St}\left( \mathbf{x}\right) =-f\frac{\exp \left( -s\left\vert \mathbf{x}%
\right\vert \right) }{\left\vert \mathbf{x}\right\vert }  \label{VSt1}
\end{equation}%
is applied, where the strong coupling strength $f=0.08\hbar c$ \cite{Bjorken}
and $1/s$ is the range of the strong interaction. ($\hbar $ is the reduced
Planck constant, $c$ is the velocity of light and $e$ is the elementary
charge.)

According to the standard perturbation theory of quantum mechanics the
transition probability per unit time $\left( W_{fi}\right) $ of this second
order process can be written as%
\begin{equation}
W_{fi}=\frac{2\pi }{\hbar }\sum_{f}\left\vert T_{fi}\right\vert ^{2}\delta
(E_{f}-E_{i}-\Delta )  \label{Wfie}
\end{equation}%
with 
\begin{equation}
T_{fi}=\sum_{\mu }\frac{V_{f\mu }^{St}V_{\mu i}^{Cb}}{\Delta E_{\mu i}}.
\label{Tif}
\end{equation}%
Here $V_{\mu i}^{Cb}$ \ is the matrix element of the Coulomb potential
between the initial and intermediate states and $V_{f\mu }^{St}$ is the
matrix element of the potential of the strong interaction between
intermediate and final states, furthermore%
\begin{equation}
\Delta E_{\mu i}=E_{\mu }-E_{i}-\Delta _{i\mu }.  \label{DeltaEmui}
\end{equation}%
$E_{i}$, $E_{\mu }$ and $E_{f}$ are the kinetic energies in the initial,
intermediate and final states, respectively, $\Delta $ is the reaction
energy,~and $\Delta _{i\mu }$ is the difference between the rest energies of
the initial $\left( _{Z}^{A_{1}}X\right) $ and intermediate $\left(
_{Z}^{A_{1}-1}X\text{ and }n\right) $ states.%
\begin{equation}
\Delta =\Delta _{-}+\Delta _{+},\text{ }\Delta _{i\mu }=\Delta _{-}-\Delta
_{n}  \label{Delta}
\end{equation}%
with 
\begin{equation}
\Delta _{-}=\Delta _{A_{1}}-\Delta _{A_{1}-1}\text{ and }\Delta _{+}=\Delta
_{A_{2}}-\Delta _{A_{2}+1}.  \label{Delta-}
\end{equation}%
$\Delta _{A_{1}}$, $\Delta _{A_{1}-1}$, $\Delta _{A_{2}}$, $\Delta
_{A_{2}+1} $ and $\Delta _{n}$\ are the energy excesses of the neutral atoms
of mass numbers $A_{1}$, $A_{1}-1$, $A_{2}$, $A_{2}+1$ and the neutron,
respectively. \cite{Shir}. The sum of initial kinetic energies $\left(
E_{i}\right) $ is neglected in the energy Dirac-delta $\delta
(E_{f}-E_{i}-\Delta )$ and $\Delta E_{\mu i}$ further on.

Particle $e$ is an electron, particle $1$ is initially the nucleus $%
_{Z}^{A_{1}}X$ and finally $_{Z}^{A_{1}-1}X$, particle $2$ is initially the
nucleus $_{Z}^{A_{2}}X$ and finally $_{Z}^{A_{2}+1}X$. 
\begin{equation}
E_{f}=E_{fe}\left( \mathbf{k}_{fe}\right) +E_{f1}\left( \mathbf{k}%
_{1}\right) +E_{f2}\left( \mathbf{k}_{2}\right) ,  \label{Ef}
\end{equation}

\begin{equation}
E_{\mu }=E_{fe}\left( \mathbf{k}_{fe}\right) +E_{\mu 1}\left( \mathbf{k}%
_{1}\right) +E_{n}\left( \mathbf{k}_{n}\right) ,  \label{Em}
\end{equation}%
where 
\begin{equation}
E_{fj}\left( \mathbf{k}_{j}\right) =\frac{\hbar ^{2}\mathbf{k}_{j}^{2}}{%
2m_{j}}  \label{Efj}
\end{equation}%
is the kinetic energy, $\mathbf{k}_{fj}\equiv \mathbf{k}_{j}$ is the wave
vector and $m_{j}$ is the rest mass of particle $j$ in the final state $%
\left( j=1,2\right) $. 
\begin{equation}
E_{n}\left( \mathbf{k}_{n}\right) =\frac{\hbar ^{2}\mathbf{k}_{n}^{2}}{2m_{n}%
}  \label{En}
\end{equation}%
is the kinetic energy, $\mathbf{k}_{n}$ is the wave vector and $m_{n}$ is
the rest mass of the neutron in the intermediate state. $E_{\mu 1}\left( 
\mathbf{k}_{1}\right) $ is the kinetic energy of the first particle in the
intermediate state, and $E_{\mu 1}\left( \mathbf{k}_{1}\right) =E_{f1}\left( 
\mathbf{k}_{1}\right) $. The kinetic energy of the electron in the initial
and final state 
\begin{equation}
E_{ie}=\frac{\hbar ^{2}\mathbf{k}_{ie}^{2}}{2m_{e}}\text{ and }E_{fe}=\frac{%
\hbar ^{2}\mathbf{k}_{fe}^{2}}{2m_{e}}  \label{E1f}
\end{equation}%
with $\mathbf{k}_{ie}$ and $\mathbf{k}_{fe}$ denoting the wave vector of the
electron in the initial and final state. The initial wave vectors $\mathbf{k}%
_{i1}$ and $\mathbf{k}_{i2}$ of particles $1$ and $2$ are neglected. The
initial, intermediate and final states are determined in Appendix A., the $%
V_{\mu i}^{Cb}$, $V_{f\mu }^{St}$ matrix-elements are calculated in Appendix
B. and the transition probability per unit time is calculated in Appendix
C.. Appendix D. is devoted to the approximations, identities and relations
which are used in the calculation of the cross section.

\section{Cross section and yield of events of electron assisted neutron
exchange process}

\subsection{ Cross section of electron assisted neutron exchange process}

The cross section $\sigma $ of the process can be obtained from the
transition probability per unit time $\left( \ref{Wfi22}\right) $ dividing
it by the flux $v_{e}/V$ \ of the incoming electron where $v_{e}$ is the
velocity of the electron. 
\begin{eqnarray}
\sigma  &=&\int \frac{c}{v_{e}}\frac{\alpha _{f}^{2}\hbar
cZ^{2}\sum_{l_{2}=-m_{2}}^{l_{2}=m_{2}}\left\vert F_{2}\left( \mathbf{k}%
_{2}\right) \right\vert ^{2}}{\pi ^{3}v_{c}\left( \left\vert \mathbf{k}_{1}+%
\mathbf{k}_{2}\right\vert ^{2}+\lambda ^{2}\right) ^{2}\left( \Delta E_{\mu
i}\right) _{\mathbf{k}_{n}=\mathbf{k}_{2}}^{2}}  \label{sigma} \\
&&\times \frac{F_{e}(E_{ie})}{F_{e}(E_{f1})}\left\langle \left\vert
F_{1}\left( \mathbf{k}_{2}\right) \right\vert ^{2}\right\rangle
A_{2}^{2}r_{A_{2}}\delta (E_{f}-\Delta )d^{3}k_{1}d^{3}k_{2},  \notag
\end{eqnarray}%
where $v_{c}$ is the volume of elementary cell in the solid, $r_{A_{2}}$ is
the relative natural abundance of atoms $_{Z}^{A_{2}}X$, 
\begin{equation}
F_{1}\left( \mathbf{k}_{2}\right) =\int \Phi _{i1}\left( \mathbf{r}%
_{n1}\right) e^{-i\mathbf{k}_{2}\frac{A_{1}}{A_{1}-1}\cdot \mathbf{r}%
_{n1}}d^{3}r_{n1},  \label{F1kalk2}
\end{equation}%
\begin{equation}
\left\langle \left\vert F_{1}\left( \mathbf{k}_{2}\right) \right\vert
^{2}\right\rangle =\frac{1}{2l_{1}+1}\sum_{l_{1}=-m_{1}}^{l_{1}=m_{1}}\left%
\vert F_{1}\left( \mathbf{k}_{2}\right) \right\vert ^{2}  \label{F1av}
\end{equation}%
and%
\begin{eqnarray}
F_{2}\left( \mathbf{k}_{2}\right)  &=&\int \Phi _{f2}^{\ast }\left( \mathbf{r%
}_{n2}\right) e^{i\mathbf{k}_{2}\cdot \mathbf{r}_{n2}}\times   \label{F2k2}
\\
&&\times \left( -f\frac{\exp (-s\frac{A_{2}+1}{A_{2}}r_{n2}}{\frac{A_{2}+1}{%
A_{2}}r_{n2}}\right) d^{3}r_{n2}.  \notag
\end{eqnarray}%
Here $\Phi _{i1}$ and $\Phi _{f2}$ are the initial and final bound neutron
states. The details of the cross section calculation (see in Appendix D.)
result that the $k_{2}\simeq k_{0}=\sqrt{2\mu _{12}\Delta }/\hbar $
substitution may be used in calculating $F_{1}$ and $F_{2}$ in $\sigma $,
where $\mu _{12}=m_{0}$ $\left[ \left( A_{1}-1\right) \left( A_{2}+1\right) %
\right] /\left( A_{1}+A_{2}\right) $.

Evaluating $\left( \ref{sigma}\right) $ first the Weisskopf approximation is
applied, i.e. for the initial and final bound neutron states we take $\Phi
_{W}\left( \mathbf{r}_{nj}\right) =\phi \left( r_{nj}\right)
Y_{l_{j}m_{j}}\left( \Omega _{j}\right) ,$ \ $j=1,2$ where $%
Y_{l_{j}m_{j}}\left( \Omega _{j}\right) $ is a spherical harmonics and $\phi
_{jW}\left( r_{nj}\right) =\sqrt{3/R_{j}^{3}},$ \ $j=1,2$ if $\left\vert 
\mathbf{r}_{nj}\right\vert \leq R_{j}$ and $\phi _{jW}\left( r_{nj}\right) =0
$ for $\left\vert \mathbf{r}_{nj}\right\vert >R_{j}$, where $%
R_{j}=r_{0}A_{j}^{1/3}$is the radius of a nucleus of nucleon number $A_{j}$
with $r_{0}=1.2\times 10^{-13}$ $cm$. We apply the $A_{1}\simeq A_{2}\simeq
A_{1}-1\simeq A_{2}+1=A$ approximation further on. Calculating $F_{1}\left( 
\mathbf{k}_{0}\right) $ and $F_{2}\left( \mathbf{k}_{0}\right) $ the long
wavelength approximations (LWA) ($\exp \left( -i\mathbf{k}_{0}\cdot \mathbf{r%
}_{n1}\right) =1$ and $\exp \left( i\mathbf{k}_{0}\cdot \mathbf{r}%
_{n2}\right) =1$) are also used with $s=1/r_{0}$ that result approximately 
\begin{equation}
\left\langle \left\vert F_{1}\left( \mathbf{k}_{0}\right) \right\vert
^{2}\right\rangle \sum_{l_{2}=-m_{2}}^{l_{2}=m_{2}}\left\vert F_{2}\left( 
\mathbf{k}_{0}\right) \right\vert ^{2}=16\pi ^{2}r_{0}^{4}f^{2}\left(
2l_{2}+1\right) .  \label{F1K2av}
\end{equation}%
Using the results of Appendix D., the $E_{f1}=\Delta /2$ relation and if $%
E_{e}<0.1$ $MeV$ (i.e. if $F_{e}(E_{ie})=\left\vert 2\pi \eta _{e}\left(
E_{ie}\right) \right\vert =2\pi Z\alpha _{f}\sqrt{m_{e}c^{2}/2E_{ie}}$) then
the cross section in the Weisskopf-LWA approximation reads as 
\begin{equation}
\sigma _{W}=\frac{C_{W0}\left( 2l_{2}+1\right) }{\left[ 1+\frac{2\left(
\Delta _{n}-\Delta _{-}\right) }{A\Delta }\right] ^{2}}\frac{r_{A_{2}}}{%
F_{e}(\Delta /2)}\frac{A^{3/2}Z^{2}}{\Delta ^{3/2}E_{ie}}  \label{sigma2}
\end{equation}%
with $C_{W0}=2^{9}\pi ^{3}\alpha _{f}^{3}\left( 0.08\right)
^{2}a_{B}r_{0}\left( \frac{r_{0}}{d}\right) ^{3}\left( m_{0}c^{2}\right)
^{3/2}m_{e}c^{2}$. Here $a_{B}$ is the Bohr-radius, the relation $c/v_{e}=%
\sqrt{m_{e}c^{2}/\left( 2E_{ie}\right) }$ with $E_{ie}$ the kinetic energy
of the ingoing electrons is also applied and $d=3.52\times 10^{-8}$ $cm$ ($Ni
$ lattice) and $d=3.89\times 10^{-8}$ $cm$ ($Pd$ lattice). $F_{e}(\Delta /2)$
is determined by $\left( \ref{Fe}\right) $ and $\left( \ref{etae}\right) $.
The subscript $W$ refers to the Weisskopf-LWA approximation and in $\left( %
\ref{sigma2}\right) $ the quantities $\Delta $ and $E_{ie}$ have to be
substituted in $MeV$ units. $C_{W0}\left( Ni\right) =1.4\times 10^{-14}$ $%
MeV^{5/2}b$ and $C_{W0}\left( Pd\right) =1.1\times 10^{-14}$ $MeV^{5/2}b$.

We have calculated $\sum_{l_{2}=-m_{2}}^{l_{2}=m_{2}}\left\vert F_{2}\left( 
\mathbf{k}_{0}\right) \right\vert ^{2}$, $\left\langle \left\vert
F_{1}\left( \mathbf{k}_{0}\right) \right\vert ^{2}\right\rangle $ and the
cross section in the single particle shell model with isotropic harmonic
oscillator potential and without the long wavelength approximation (see
Appendix E.). We introduce the ratio 
\begin{equation}
\eta =\frac{\left\langle \left\vert F_{1}\left( \mathbf{k}_{0}\right)
\right\vert ^{2}\right\rangle
_{Sh}\sum_{l_{2}=-m_{2}}^{l_{2}=m_{2}}\left\vert F_{2}\left( \mathbf{k}%
_{0}\right) \right\vert _{Sh}^{2}}{\left\langle \left\vert F_{1}\left( 
\mathbf{k}_{0}\right) \right\vert ^{2}\right\rangle
_{W}\sum_{l_{2}=-m_{2}}^{l_{2}=m_{2}}\left\vert F_{2}\left( \mathbf{k}%
_{0}\right) \right\vert _{W}^{2}}.  \label{etha}
\end{equation}%
(The subscript $Sh$ refers to the shell model.) With the aid of $\eta \equiv
\eta _{l_{1},n_{1},l_{2},n_{2}}\left( A_{1},A_{2}\right) $ given by $\left( %
\ref{etha2}\right) $ (see Appendix E.) the cross section $\sigma _{Sh}$
calculated in the shell model can be written as 
\begin{equation}
\sigma _{Sh}=\eta _{l_{1},n_{1},l_{2},n_{2}}\left( A_{1},A_{2}\right) \sigma
_{W}.  \label{sigmaSh}
\end{equation}

\subsection{Yield of events of electron assisted neutron exchange process}

The yield $dN/dt$ of events of electron assisted neutron exchange process $%
A_{1},A_{2}\rightarrow A_{1}-1,A_{2}+1$ can be written as 
\begin{equation}
\frac{dN}{dt}=N_{t}N_{ni}\sigma \Phi ,  \label{rate1}
\end{equation}%
where $\sigma =\left\{ \sigma _{W}\text{ or }\sigma _{Sh}\right\} $, $\Phi $
is the flux of electrons, $N_{t}$ is the number of target particles, i.e.
the number $N_{A_{1}}$ of irradiated atoms of mass number $A_{1}$ in the
metal. The contribution of $N_{ni}$ neutrons in each nucleus $_{Z}^{A_{1}}X$
is also taken into account. $N_{ni}$ is the number of neutrons in the
uppermost energy level of the initial nucleus $_{Z}^{A_{1}}X$. If $F$ and $D$
are the irradiated surface and the width of the sample, respectively, then
the number of elementary cells $N_{c}$ in the sample is $%
N_{c}=FD/v_{c}=4FD/d^{3}$ in the case of $Ni$ and $Pd$, and the number of
atoms in the elementary cell is $2r_{A_{1}}$ with $r_{A_{1}}$ the relative
natural abundance of atoms $_{Z}^{A_{1}}X$ thus the number $N_{t}$ of target
atoms of mass number $A_{1}$ in the process is 
\begin{equation}
N_{t}=\frac{8}{d^{3}}r_{A_{1}}FD.  \label{NA1}
\end{equation}

The wave numbers and energies of the two outgoing heavy particles are
approximately $\mathbf{k}_{1}=-\mathbf{k}_{2}$,%
\begin{equation}
E_{1}=\frac{A_{2}+1}{A_{1}+A_{2}}\Delta \text{ and }E_{2}=\frac{A_{1}-1}{%
A_{1}+A_{2}}\Delta .  \label{E1}
\end{equation}

\section{Numerical data of electron assisted neutron exchange processes in $%
Ni$ and $Pd$}

\begin{table}[tbp]
\tabskip=8pt 
\centerline {\vbox{\halign{\strut $#$\hfil&\hfil$#$\hfil&\hfil$#$
\hfil&\hfil$#$\hfil&\hfil$#$\hfil&\hfil$#$\hfil&\hfil$#$\cr
\noalign{\hrule\vskip2pt\hrule\vskip2pt}
A &58 &60 &61 &62 &64 \cr
\Delta_{-} &-4.147 &-3.317 &0.251 &-2.526 &-1.587 \cr
\Delta_{+} &0.928 &-0.251 &2.526 &-1.234 &-1.973 \cr
r_{A} &0.68077 &0.26223 &0.0114 &0.03634 &0.00926 \cr
\noalign{\vskip2pt\hrule\vskip2pt\hrule}}}}
\caption{Numerical data of the $\text{ }e+\text{ }_{28}^{A_{1}}Ni+\text{ }%
_{28}^{A_{2}}Ni\rightarrow e^{\prime }+\text{ }_{28}^{A_{1}-1}Ni+\text{ }%
_{28}^{A_{2}+1}Ni+\Delta $ reaction. The reaction is energetically allowed
if $\Delta =\Delta _{-}(A_{1})+\Delta _{+}(A_{2})>0$ holds. $A$ is the mass
number, $r_{A}$ is the relative natural abundance, $\Delta _{-}(A)=\Delta
_{A}-\Delta _{A-1}$ and $\Delta _{+}(A)=\Delta _{A}-\Delta _{A+1}$ are given
in $MeV$ units.}
\label{Table1}
\end{table}

\begin{table}[tbp]
\tabskip=8pt 
\centerline {\vbox{\halign{\strut $#$\hfil&\hfil$#$\hfil&\hfil$#$
\hfil&\hfil$#$\hfil&\hfil$#$\hfil&\hfil$#$\hfil&\hfil$#$\cr
\noalign{\hrule\vskip2pt\hrule\vskip2pt}
A &102 &104 &105 &106 &108 &110 \cr
\Delta_{-} &-2.497 &-1.912 &0.978 &-1.491 &-1.149 &-0.747 \cr
\Delta_{+} &-0.446 &-0.978 &1.491 &-1.533 &-1.918 &-2.320 \cr
r_{A} &0.0102 &0.1114 &0.2233 &0.2733 &0.2646 &0.1172 \cr
\noalign{\vskip2pt\hrule\vskip2pt\hrule}}}}
\caption{Numerical data of the $\text{ }e+\
_{46}^{A_{1}}Pd+_{46}^{A_{2}}Pd\rightarrow e^{\prime }+\text{ }%
_{46}^{A_{1}-1}Pd+_{46}^{A_{2}+1}Pd+\Delta$ reaction. The reaction is
energetically allowed if $\Delta=\Delta_{-}(A_{1})+\Delta_{+}(A_{2})>0$
holds. $A$ is the mass number, $r_{A}$ is the relative natural abundance, $%
\Delta_{-}(A)=\Delta_{A}-\Delta_{A-1} $ and $\Delta_{+}(A)=\Delta_{A}-%
\Delta_{A+1} $ are given in $MeV$ units.}
\label{Table2}
\end{table}

\begin{table}[tbp]
\tabskip=8pt 
\centerline {\vbox{\halign{\strut $#$\hfil &\hfil$#$\hfil&\hfil$#$
\hfil&\hfil$#$\hfil\cr
\noalign{\hrule\vskip2pt\hrule\vskip2pt}
 A_{1}\rightarrow A_{1}-1&A_{2}\rightarrow A_{2}+1&\Delta($MeV$)&\eta \cr
\noalign{\vskip2pt\hrule\vskip2pt}
61 \rightarrow 60 &58 \rightarrow 59 & 1.179 &7.02\times10^{-3}\cr
61 \rightarrow 60 &61 \rightarrow 62 & 2.777 &2.42\times10^{-8}\cr
64 \rightarrow 63 &61 \rightarrow 62 & 0.939 &2.08\times10^{-4}\cr
\noalign{\vskip2pt\hrule\vskip2pt\hrule}}}}
\caption{The values of the quantities $\protect\eta$ and $\Delta =\Delta
_{-}(A_{1})+\Delta _{+}(A_{2})>0$, the later in $MeV $ units, of the $\text{ 
}e+\text{ }_{28}^{A_{1}}Ni+\text{ }_{28}^{A_{2}}Ni\rightarrow e^{\prime }+%
\text{ }_{28}^{A_{1}-1}Ni+\text{ }_{28}^{A_{2}+1}Ni+\Delta $ reaction. The $%
\Delta _{-}(A_{1})$ and $\Delta _{+}(A_{2})$ values can be found in Table I.
For the definition of $\protect\eta $ see $\left( \protect\ref{etha}\right)$
and $\left( \protect\ref{etha2}\right)$.}
\label{Table3}
\end{table}

\begin{table}[tbp]
\tabskip=8pt 
\centerline {\vbox{\halign{\strut $#$\hfil &\hfil$#$\hfil&\hfil$#$
\hfil&\hfil$#$\hfil\cr
\noalign{\hrule\vskip2pt\hrule\vskip2pt}
 A_{1}\rightarrow A_{1}-1&A_{2}\rightarrow A_{2}+1&\Delta($MeV$)&\eta \cr
\noalign{\vskip2pt\hrule\vskip2pt}
105 \rightarrow 104 &102 \rightarrow 103 & 0.532 & 1.84\times10^{-4}\cr
105 \rightarrow 104 &105 \rightarrow 106 & 2.469 & 8.88\times10^{-11}\cr
108 \rightarrow 107 &105 \rightarrow 106 & 0.342 & 2.82\times10^{-3}\cr
\noalign{\vskip2pt\hrule\vskip2pt\hrule}}}}
\caption{The values of the quantities $\protect\eta$ and $\Delta =\Delta
_{-}(A_{1})+\Delta _{+}(A_{2})>0$, the later in $MeV $ units, of the $\text{ 
}e+\ _{46}^{A_{1}}Pd+_{46}^{A_{2}}Pd\rightarrow e^{\prime }+\text{ }%
_{46}^{A_{1}-1}Pd+_{46}^{A_{2}+1}Pd+\Delta $ reaction. The $\Delta
_{-}(A_{1})$ and $\Delta _{+}(A_{2})$ values can be found in Table II. For
the definition of $\protect\eta$ see $\left( \protect\ref{etha}\right)$ and $%
\left( \protect\ref{etha2}\right)$.}
\label{Table4}
\end{table}

As a first example we take $Ni$ as target material. In this case the
possible processes are%
\begin{equation}
\text{ }e+\text{ }_{28}^{A_{1}}Ni+\text{ }_{28}^{A_{2}}Ni\rightarrow
e^{\prime }+\text{ }_{28}^{A_{1}-1}Ni+\text{ }_{28}^{A_{2}+1}Ni+\Delta .
\label{NiAp}
\end{equation}%
Tables I. and III. contain the relevant data for reaction $\left( \ref{NiAp}%
\right) $. Describing neutrons in the uppermost energy level of $_{28}^{A}Ni$
isotopes we used $1p$ shell model states in the cases of $A=58-60$ and $0f$
shell model states in the cases of $A=61-64$.

Another interesting target material is $Pd$ in which the electron assisted
neutron exchange processes are the 
\begin{equation}
\text{ }e+\ _{46}^{A_{1}}Pd+\text{ }_{46}^{A_{2}}Pd\rightarrow e^{\prime }+%
\text{ }_{46}^{A_{1}-1}Pd+\text{ }_{46}^{A_{2}+1}Pd+\Delta  \label{PdAp}
\end{equation}%
reactions. The relevant data can be found in Tables II. and IV.. Describing
neutrons in the uppermost energy level of $_{46}^{A}Pd$ isotopes we used $0g$
shell model states in the cases of $A=102-104$ and $1d$ shell model states
in the cases of $A=105-108$. The nuclear data to the Tables are taken from 
\cite{Shir}. One can see from Tables III. and IV. that in both cases three
possible pairs of isotopes exist which are energetically allowed (for which $%
\Delta >0$) and their rates differ in the factor $\left( 2l_{2}+1\right)
N_{ni}\eta _{l_{1},n_{1},l_{2},n_{2}}\left( A_{1},A_{2}\right)
r_{A_{1}}r_{A_{2}}\Delta ^{-3/2}$ only. The $\eta \equiv \eta
_{l_{1},n_{1},l_{2},n_{2}}\left( A_{1},A_{2}\right) $ values of $Ni$ and $Pd$
can also be found in Tables III. and IV., respectively. The results of
numerical investigation of $\left( 2l_{2}+1\right) N_{ni}\eta
_{l_{1},n_{1},l_{2},n_{2}}\left( A_{1},A_{2}\right) r_{A_{1}}r_{A_{2}}\Delta
^{-3/2}$ shows that the $61\rightarrow 60,58\rightarrow 59$ and the $%
108\rightarrow 107,105\rightarrow 106$ reactions are the dominant among the
processes in $Ni$ and $Pd$, respectively.

\section{Summary}

The electron assisted neutron exchange process is discussed. The transition
probability per unit time and the cross section of the process are
determined in those cases when the electron takes off negligible energy. The
electron assisted neutron exchange processes are investigated numerically in 
$Ni$ and $Pd$. In the case of $Ni$ it is found that the 
\begin{equation}
\text{ }e+\text{ }_{28}^{61}Ni+\text{ }_{28}^{58}Ni\rightarrow e^{\prime }+%
\text{ }_{28}^{60}Ni+\text{ }_{28}^{59}Ni+1.179\text{ }MeV  \label{Nilead}
\end{equation}%
process of $\sigma _{Sh}=0.088/E_{ie}$ $\mu b$ with $E_{ie}$ in $MeV$ is
leading. In this case the $_{28}^{60}Ni$ and the $_{28}^{59}Ni$ isotopes
take away $0.585$ $MeV$ and $0.594$ $MeV$, respectively. In the case of $Pd$
the 
\begin{equation}
e+\text{ }_{46}^{108}Pd+\text{ }_{46}^{105}Pd\rightarrow e^{\prime }+\text{ }%
_{46}^{107}Pd+\text{ }_{46}^{106}Pd+0.342\text{ }MeV  \label{Pdlead}
\end{equation}%
reaction of $\sigma _{Sh}=0.26/E_{ie}$ $\mu b$ with $E_{ie}$ in $MeV$ is
found to be the leading one. In this case the $_{46}^{107}Pd$ and the $%
_{46}^{106}Pd$ isotopes take away $0.170$ $MeV$ and $0.172$ $MeV$,
respectively.

There are many other materials which may be suitable for hosting electron
assisted neutron exchange process some of which we list in the following. We
deal with period 4 of transition metals. (The metal targets are advantageous
from experimental point of view since in the case of metals it is easy to
avoid the charging of the sample.) The $Cr$, $Fe$ and $V$ have body centered
cubic crystal lattice. In the case of $V$ there is one and in the case of $%
Cr $ and $Fe$ there are two energetically allowed electron assisted neutron
exchange processes. $Ti$ and $Zn$ have closepacked hexagonal crystal
structure. In the case of\ $Ti$ there are five and in the case of $Zn$ there
are three energetically allowed electron assisted neutron exchange
processes. Since the natural abundance of $_{25}^{45}Sc$, $_{25}^{55}Mn$ and 
$_{27}^{59}Co$ equals unity in the case of these materials there is no
chance of the electron assisted neutron exchange process. Although $Cu$ has
two natural isotopes, their electron assisted neutron exchange processes are
energetically forbidden.

\section{Appendix}

\subsection{Initial, intermediate and final states of the process}

Let $\Psi _{i}$, $\Psi _{\mu }$ and $\Psi _{f}$ denote the space dependent
parts of initial, intermediate and final states, respectively. The initial
state has the form 
\begin{equation}
\Psi _{i}(\mathbf{x}_{e},\mathbf{x}_{1},\mathbf{x}_{n1},\mathbf{x}_{2})=\psi
_{ie}\left( \mathbf{x}_{e}\right) \psi _{i1n}(\mathbf{x}_{1},\mathbf{x}%
_{n1})\psi _{i2}(\mathbf{x}_{2}),  \label{Pszii}
\end{equation}%
where 
\begin{equation}
\psi _{ie}\left( \mathbf{x}_{e}\right) =V^{-1/2}e^{\left( i\mathbf{k}%
_{ie}\cdot \mathbf{x}_{e}\right) }\text{ and }\psi _{i2}(\mathbf{x}%
_{2})=V^{-1/2}e^{\left( i\mathbf{k}_{i2}\cdot \mathbf{x}_{2}\right) }
\label{psziei}
\end{equation}%
are the initial state of the electron and the nucleus $_{Z}^{A_{2}}X$, and $%
\psi _{i1n}(\mathbf{x}_{1},\mathbf{x}_{n1})$ is the initial state of the
neutron and the initial $A_{1}-1$ nucleon of the nucleus $_{Z}^{A_{1}}X$. $%
\mathbf{x}_{e}$, $\mathbf{x}_{1},\mathbf{x}_{n1}$ and $\mathbf{x}_{2}$ are
the coordinates of the electron, the center of mass of the initial $A_{1}-1$
nucleon, the neutron and the nucleus $_{Z}^{A_{2}}X$, respectively. $\mathbf{%
k}_{ie}$ and $\mathbf{k}_{i2}$ are the initial wave vectors of the electron
and the nucleus $_{Z}^{A_{2}}X$ and $V$ is the volume of normalization. The
initial state $\psi _{i1n}(\mathbf{x}_{1},\mathbf{x}_{n1})$ of the neutron
and the initial $A_{1}-1$ nucleon may be given in the variables $\mathbf{R}%
_{1}$, $\mathbf{r}_{n1}$ 
\begin{equation}
\psi _{i1n}(\mathbf{R}_{1},\mathbf{r}_{n1})=V^{-1/2}\exp (i\mathbf{k}%
_{i1}\cdot \mathbf{R}_{1})\Phi _{i1}\left( \mathbf{r}_{n1}\right)
\label{pszii1}
\end{equation}%
where $\mathbf{R}_{1}$ is the center of mass coordinate of the nucleus $%
_{Z}^{A_{1}}X$ and $\mathbf{r}_{n1}$ is the relative coordinate of one of
its neutrons. $\mathbf{R}_{1}$ and $\mathbf{r}_{n1}$are determined by the
usual $\mathbf{x}_{n1}=\mathbf{R}_{1}+\mathbf{r}_{n1}$ and $\mathbf{R}_{1}=%
\left[ \left( A_{1}-1\right) \mathbf{x}_{1}+\mathbf{x}_{n1}\right] /A_{1}$
relations where $\mathbf{x}_{n1}$ and $\mathbf{x}_{1}$ are the coordinates
of the neutron and of the center of mass of the initial $A_{1}-1$ nucleon,
respectively. The inverse formula for $\mathbf{x}_{1}$ is $\mathbf{x}_{1}=%
\mathbf{R}_{1}-\mathbf{r}_{n1}/\left( A_{1}-1\right) $. In $\left( \ref%
{pszii1}\right) $ the $\Phi _{i1}\left( \mathbf{r}_{n1}\right) $ is the wave
function of the neutron in the initial bound state of nucleus $_{Z}^{A_{1}}X$%
, $\mathbf{k}_{i1}$is the initial wave vector of nucleus $_{Z}^{A_{1}}X$.

The intermediate state has the form 
\begin{equation}
\Psi _{\mu }(\mathbf{x}_{e},\mathbf{x}_{1},\mathbf{x}_{n1},\mathbf{x}%
_{2})=\psi _{fe}\left( \mathbf{x}_{e}\right) \psi _{\mu 1n}(\mathbf{x}_{1},%
\mathbf{x}_{n1})\psi _{i2}(\mathbf{x}_{2}),  \label{Pszimu}
\end{equation}%
where 
\begin{equation}
\psi _{fe}\left( \mathbf{x}_{e}\right) =V^{-1/2}e^{\left( i\mathbf{k}%
_{fe}\cdot \mathbf{x}_{e}\right) }  \label{pszief}
\end{equation}%
with $\mathbf{k}_{fe}$ the wave vector of the electron in the final state
and $\psi _{i2}(\mathbf{x}_{2})$ is given in $\left( \ref{psziei}\right) $.
The state $\psi _{\mu 1n}(\mathbf{x}_{1},\mathbf{x}_{n1})$ is the product of
two plane waves $\psi _{f1}(\mathbf{x}_{1})=V^{-1/2}e^{\left( i\mathbf{k}%
_{1}\cdot \mathbf{x}_{1}\right) }$ and $\psi _{n}\left( \mathbf{x}%
_{n1}\right) =V^{-1/2}e^{i\mathbf{k}_{n}\cdot \mathbf{x}_{n1}}$, which are
the final state of the nucleus $_{Z_{1}}^{A_{1}-1}X$ and the state of the
free, intermediate neutron. Thus $\psi _{\mu 1n}(\mathbf{x}_{1},\mathbf{x}%
_{n1})=V^{-1}e^{i\mathbf{k}_{1}\cdot \mathbf{x}_{1}}e^{i\mathbf{k}_{n}\cdot 
\mathbf{x}_{n1}}$ and it has the form in the coordinates $\mathbf{R}_{1}$, $%
\mathbf{r}_{n1}$%
\begin{equation}
\psi _{\mu 1n}(\mathbf{R}_{1},\mathbf{r}_{n1})=V^{-1}e^{i\left( \mathbf{k}%
_{1}+\mathbf{k}_{n}\right) \cdot \mathbf{R}_{1}}e^{i\left( \mathbf{k}_{n}-%
\frac{\mathbf{k}_{1}}{A_{1}-1}\right) \mathbf{r}_{n1}},  \label{pszimu2}
\end{equation}%
where $\mathbf{k}_{1}$ and $\mathbf{k}_{n}$ are the wave vectors of the
nucleus $_{Z}^{A_{1}-1}X$ and the neutron, respectively.

The intermediate state may have an other form%
\begin{equation}
\Psi _{\mu }(\mathbf{x}_{e},\mathbf{x}_{1},\mathbf{x}_{n1},\mathbf{x}%
_{2})=\psi _{fe}\left( \mathbf{x}_{e}\right) \psi _{f1}(\mathbf{x}_{1})\psi
_{\mu 2n}(\mathbf{x}_{n1},\mathbf{x}_{2}),  \label{Pszimu2}
\end{equation}%
where 
\begin{equation}
\psi _{\mu 2n}(\mathbf{x}_{n1},\mathbf{x}_{2})=\psi _{n}\left( \mathbf{x}%
_{n1}\right) \psi _{i2}(\mathbf{x}_{2})=V^{-1}e^{i\mathbf{k}_{n}\cdot 
\mathbf{x}_{n1}}e^{i\mathbf{k}_{i2}\cdot \mathbf{x}_{2}}  \label{pszimu3}
\end{equation}%
which can be written in the coordinates $\mathbf{r}_{n2}=\mathbf{x}_{n1}-%
\mathbf{R}_{2}$ and $\mathbf{R}_{2}=\left( A_{2}\mathbf{x}_{2}+\mathbf{x}%
_{n1}\right) /\left( A_{2}+1\right) $ as 
\begin{equation}
\psi _{\mu 2n}(\mathbf{R}_{2},\mathbf{r}_{n2})=\frac{1}{V}e^{i\left( \mathbf{%
k}_{i2}+\mathbf{k}_{n}\right) \cdot \mathbf{R}_{2}}e^{i\left( \mathbf{k}_{n}-%
\frac{\mathbf{k}_{i2}}{A_{2}}\right) \mathbf{r}_{n2}},  \label{pszimu4}
\end{equation}%
where $\mathbf{R}_{2}$ is the center of mass coordinate of the nucleus $%
_{Z}^{A_{2}+1}X$ and $\mathbf{r}_{n2}$ is the relative coordinate of the
neutron in it. In these new variables $\mathbf{x}_{2}=\mathbf{R}_{2}-\mathbf{%
r}_{n2}/A_{2}$ and $\mathbf{x}_{n1}-\mathbf{x}_{2}=\left( A_{2}+1\right) 
\mathbf{r}_{n2}/A_{2}$ which is used in the argument of $V^{St}$ (given by $%
\left( \ref{VSt1}\right) $) in calculating $V_{f\mu }^{St}$. Evaluating the
matrix elements $V_{\mu i}^{Cb}$ and $V_{f\mu }^{St}$ the forms $\left( \ref%
{pszimu2}\right) $ and $\left( \ref{pszimu4}\right) $ of $\psi _{\mu }$ are
used, respectively, and $\sum_{\mu }\rightarrow \frac{V}{\left( 2\pi \right)
^{3}}d^{3}k_{n}$ in $\left( \ref{Tif}\right) $.

The final state has the form 
\begin{equation}
\Psi _{f}(\mathbf{x}_{e},\mathbf{x}_{1},\mathbf{x}_{n1},\mathbf{x}_{2})=\psi
_{fe}\left( \mathbf{x}_{e}\right) \psi _{f1}(\mathbf{x}_{1})\psi _{f2n}(%
\mathbf{x}_{n1},\mathbf{x}_{2}),  \label{Pszif}
\end{equation}%
where $\psi _{f2n}(\mathbf{x}_{n1},\mathbf{x}_{2})$ is given in the
variables $\mathbf{R}_{2}$, $\mathbf{r}_{n2}$ as 
\begin{equation}
\psi _{f2n}(\mathbf{R}_{2},\mathbf{r}_{n2})=V^{-1/2}\exp (i\mathbf{k}%
_{2}\cdot \mathbf{R}_{2})\Phi _{f2}\left( \mathbf{r}_{n2}\right) ,
\label{pszif2}
\end{equation}%
and $\Phi _{f2}\left( \mathbf{r}_{n2}\right) $ is the bound state of the
neutron in the nucleus $_{Z}^{A_{2}+1}X$.

\subsection{Evaluation of matrix elements $V_{\protect\mu i}^{Cb}$ and $V_{f%
\protect\mu }^{St}$}

The argument of the Coulomb potential $V^{Cb}$ is $\mathbf{x}_{e}-\mathbf{x}%
_{1}$ therefore the integration with respect to the components of $\mathbf{x}%
_{2}$ may be carried out and $\int \left\vert \psi _{i2}(\mathbf{x}%
_{2})\right\vert ^{2}d^{3}x_{2}=1$. The remainder is 
\begin{eqnarray}
V_{\mu i}^{Cb} &=&\int \psi _{fe}^{\ast }\left( \mathbf{x}_{e}\right) \psi
_{\mu 1n}^{\ast }(\mathbf{x}_{1},\mathbf{x}_{n1})V^{Cb}\left( \mathbf{x}_{e}-%
\mathbf{x}_{1}\right)   \label{VCbmui} \\
&&\times \psi _{ie}\left( \mathbf{x}_{e}\right) \psi _{i1n}(\mathbf{x}_{1},%
\mathbf{x}_{n1})d^{3}x_{e}d^{3}x_{1}d^{3}x_{n1}.  \notag
\end{eqnarray}%
Making the $\mathbf{x}_{1},\mathbf{x}_{n1}\rightarrow \mathbf{R}_{1},\mathbf{%
r}_{n1}$ change in the variables, substituting the forms $\left( \ref{pszii1}%
\right) $ and $\left( \ref{pszimu2}\right) $ of $\psi _{i1n}$ and $\psi
_{\mu 1n}$, and neglecting $\mathbf{k}_{i1}$, the integrations over the
components of $\mathbf{x}_{e}$ and $\mathbf{R}_{1}$ result $V^{-1}\left(
2\pi \right) ^{3}\delta \left( \mathbf{q}+\mathbf{k}_{ie}-\mathbf{k}%
_{fe}\right) $ and $V^{-3/2}\left( 2\pi \right) ^{3}\delta \left( \mathbf{q}-%
\mathbf{k}_{1}-\mathbf{k}_{n}\right) $, respectively and the integration
over the components of $\mathbf{r}_{n1}$ produces $F_{1}\left( \mathbf{k}%
_{n}\right) $ where 
\begin{equation}
F_{1}\left( \mathbf{k}_{n}\right) =\int \Phi _{i1}\left( \mathbf{r}%
_{n1}\right) e^{-i\left( \mathbf{k}_{n}-\frac{\mathbf{k}_{1}+\mathbf{q}}{%
A_{1}-1}\right) \cdot \mathbf{r}_{n1}}d^{3}r_{n1}.  \label{Fk1kn}
\end{equation}%
Using the $\delta \left( \mathbf{q}+\mathbf{k}_{ie}-\mathbf{k}_{fe}\right) $
in carrying out the integration over the components of $\mathbf{q}$ in $%
V_{\mu i}^{Cb}$ one gets%
\begin{eqnarray}
V_{\mu i}^{Cb} &=&-\frac{\pi e^{2}Z}{2\pi ^{2}\left\vert \mathbf{k}_{fe}-%
\mathbf{k}_{ie}\right\vert ^{2}+\lambda ^{2}}\widetilde{F}_{1}\left( \mathbf{%
k}_{n}\right) \frac{\left( 2\pi \right) ^{6}}{V^{5/2}}\times 
\label{VCbmui1} \\
&&\times \sqrt{G_{S}}\delta \left( \mathbf{k}_{ie}-\mathbf{k}_{fe}-\mathbf{k}%
_{1}-\mathbf{k}_{n}\right)   \notag
\end{eqnarray}%
and%
\begin{equation}
\widetilde{F}_{1}\left( \mathbf{k}_{n}\right) =\int \Phi _{i1}\left( \mathbf{%
r}_{n1}\right) e^{-i\left( \mathbf{k}_{n}-\frac{\mathbf{k}_{1}+\mathbf{k}%
_{fe}-\mathbf{k}_{ie}}{A_{1}-1}\right) \cdot \mathbf{r}_{n1}}d^{3}r_{n1}.
\label{Fk1kn2}
\end{equation}%
For particles $e$ and $1$ (ingoing electron of charge $-e$ and initial
nucleus $_{Z}^{A_{1}}X$ of charge $Ze$) taking part in Coulomb interaction
we have used plane waves therefore the matrix element must be corrected with
the so called Sommerfeld factor \cite{Heitler} $\sqrt{G_{S}}$ where%
\begin{equation}
G_{S}=\frac{F_{e}(E_{ie})}{F_{e}(E_{f1})}.  \label{Gs}
\end{equation}

Now we deal with $V_{f\mu }^{St}$. The strong interaction works between the
neutron and the nucleons of the nucleus $_{Z}^{A_{2}}X$ therefore the
argument of $V^{St}$ is $\mathbf{x}_{n1}-\mathbf{x}_{2}$. The integrations
with respect to the components of $\ \mathbf{x}_{e}$ and $\mathbf{x}_{1}$
result $\int \left\vert \psi _{ef}(\mathbf{x}_{e})\right\vert
^{2}d^{3}x_{e}= $ $\int \left\vert \psi _{f1}(\mathbf{x}_{1})\right\vert
^{2}d^{3}x_{1}=1$. The remainder is 
\begin{equation}
V_{f\mu }^{St}=\int \psi _{f2n}^{\ast }V^{St}\left( \mathbf{x}_{n1}-\mathbf{x%
}_{2}\right) \psi _{\mu 2n}d^{3}x_{2}d^{3}x_{n1}.  \label{VStfmu}
\end{equation}%
Similarly to the above, making the $\mathbf{x}_{n1},\mathbf{x}%
_{2}\rightarrow \mathbf{R}_{2},\mathbf{r}_{n2}$ change in the variables,
substituting the forms $\left( \ref{pszimu4}\right) $ and $\left( \ref%
{pszif2}\right) $ of $\psi _{\mu 2n}$ and $\psi _{f2n}^{\ast }$ and
neglecting $\mathbf{k}_{i2}$, the integrations over the components of $%
\mathbf{R}_{2}$ result $V^{-3/2}\left( 2\pi \right) ^{3}\delta \left( 
\mathbf{k}_{n}-\mathbf{k}_{2}\right) $ and the integrations with respect to
the components of $\mathbf{r}_{n2}$ produces $F_{2}\left( \mathbf{k}%
_{n}\right) $ with%
\begin{eqnarray}
F_{2}\left( \mathbf{k}_{n}\right) &=&\int \Phi _{f2}^{\ast }\left( \mathbf{r}%
_{n2}\right) e^{i\mathbf{k}_{n}\cdot \mathbf{r}_{n2}}\times  \label{F2ki2kn}
\\
&&\times \left( -f\frac{\exp (-s\frac{A_{2}+1}{A_{2}}r_{n2}}{\frac{A_{2}+1}{%
A_{2}}r_{n2}}\right) d^{3}r_{n2},  \notag
\end{eqnarray}%
where $r_{n2}=\left\vert \mathbf{r}_{n2}\right\vert $. Taking into account
that the neutron interacts with each nucleon of the final nucleus of nucleon
number $A_{2}$ 
\begin{equation}
V_{f\mu }^{St}=\frac{\left( 2\pi \right) ^{3}}{V^{3/2}}A_{2}F_{2}\left( 
\mathbf{k}_{n}\right) \delta \left( \mathbf{k}_{n}-\mathbf{k}_{2}\right) .
\label{VStfmu2}
\end{equation}

\subsection{Transition probability per unit time of electron assisted
neutron exchange process}

Substituting the obtained forms of $V_{\mu i}^{Cb}$ and $V_{f\mu }^{St}$
(formulae $\left( \ref{VCbmui1}\right) $ and $\left( \ref{VStfmu2}\right) $)
into $\left( \ref{Tif}\right) $ and using the correspondence $\sum_{\mu
}\rightarrow \frac{V}{\left( 2\pi \right) ^{3}}d^{3}k_{n}$ and the $\delta
\left( \mathbf{k}_{n}-\mathbf{k}_{2}\right) $ in the integration over the
components of $\mathbf{k}_{n}$ one gets%
\begin{eqnarray}
T_{fi} &=&-\frac{e^{2}ZA_{2}\widetilde{F}_{1}\left( \mathbf{k}_{2}\right)
F_{2}\left( \mathbf{k}_{2}\right) \sqrt{\frac{F_{e}(E_{ie})}{F_{e}(E_{f1})}}%
}{2\pi ^{2}\left\vert \mathbf{k}_{fe}-\mathbf{k}_{ie}\right\vert
^{2}+\lambda ^{2}}\times   \label{Tfi22} \\
&&\times \frac{\left( 2\pi \right) ^{6}}{V^{3}}\frac{\delta \left( \mathbf{k}%
_{1}+\mathbf{k}_{2}+\mathbf{k}_{fe}-\mathbf{k}_{ie}\right) }{\left( \Delta
E_{\mu i}\right) _{\mathbf{k}_{n}=\mathbf{k}_{2}}},  \notag
\end{eqnarray}%
where%
\begin{equation}
\widetilde{F}_{1}\left( \mathbf{k}_{2}\right) =\int \Phi _{i1}\left( \mathbf{%
r}_{n1}\right) e^{-i\left( \mathbf{k}_{2}-\frac{\mathbf{k}_{1}+\mathbf{k}%
_{fe}-\mathbf{k}_{ie}}{A_{1}-1}\right) \cdot \mathbf{r}_{n1}}d^{3}r_{n1}
\label{F1k2}
\end{equation}%
and $F_{2}\left( \mathbf{k}_{2}\right) $ is determined by $\left( \ref{F2k2}%
\right) $. Here $\Phi _{i1}$ and $\Phi _{f2}$ in $\left( \ref{F2k2}\right) $
are the initial and final bound neutron states. Substituting the above into $%
\left( \ref{Wfie}\right) $, using the identities $\left[ \delta \left( 
\mathbf{k}_{1}+\mathbf{k}_{2}+\mathbf{k}_{fe}-\mathbf{k}_{ie}\right) \right]
^{2}=\delta \left( \mathbf{k}_{1}+\mathbf{k}_{2}+\mathbf{k}_{fe}-\mathbf{k}%
_{ie}\right) \delta \left( \mathbf{0}\right) $ and $\left( 2\pi \right)
^{3}\delta \left( \mathbf{0}\right) =V$, the $\sum_{f}\rightarrow
\sum_{m_{2}}\int \left[ V/\left( 2\pi \right) ^{3}\right]
^{3}d^{3}k_{1}d^{3}k_{2}d^{3}k_{fe}$ correspondence, averaging over the
quantum number $m_{1\text{ }}$and integrating over the components of $%
\mathbf{k}_{fe}$ (which gives $\mathbf{k}_{fe}=-\mathbf{k}_{1}-\mathbf{k}%
_{2}+\mathbf{k}_{ie}$) one obtains%
\begin{eqnarray}
W_{fi} &=&\int \frac{\alpha _{f}^{2}\hbar
c^{2}Z^{2}\sum_{l_{2}=-m_{2}}^{l_{2}=m_{2}}\left\vert F_{2}\left( \mathbf{k}%
_{2}\right) \right\vert ^{2}}{\pi ^{3}v_{c}V\left( \left\vert \mathbf{k}_{1}+%
\mathbf{k}_{2}\right\vert ^{2}+\lambda ^{2}\right) ^{2}\left( \Delta E_{\mu
i}\right) _{\mathbf{k}_{n}=\mathbf{k}_{2}}^{2}}  \label{Wfi22} \\
&&\times \left\langle \left\vert F_{1}\left( \mathbf{k}_{2}\right)
\right\vert ^{2}\right\rangle \frac{F_{e}(E_{ie})}{F_{e}(E_{f1})}%
A_{2}^{2}r_{A_{2}}\delta (E_{f}-\Delta )d^{3}k_{1}d^{3}k_{2},  \notag
\end{eqnarray}%
where $A_{1}$, $A_{2}$ are the initial atomic masses, $l_{1},m_{1}$ and $%
l_{2},m_{2}$ are the orbit and its projection quantum numbers of the neutron
in its initial and final state. For $F_{1}\left( \mathbf{k}_{2}\right) $, $%
\left\langle \left\vert F_{1}\left( \mathbf{k}_{2}\right) \right\vert
^{2}\right\rangle $ and $F_{2}\left( \mathbf{k}_{2}\right) $ see $\left( \ref%
{F1kalk2}\right) $, $\left( \ref{F1av}\right) $ and $\left( \ref{F2k2}%
\right) $. Taking into account the effect of the number of atoms of atomic
number $A_{2}$ in the solid target the calculation is similar to the
calculation of e.g. the coherent neutron scattering \cite{Kittel} and the $%
\left\vert T_{fi}\right\vert ^{2}$ must be multiplied by $N_{L}$ which is
the number of atomic sites in the crystal and by $r_{A_{2}}$ which is the
relative natural abundance of atoms $_{Z}^{A_{2}}X$. We have used $%
N_{L}/V=2/v_{c}$ with $v_{c}$ the volume of the elementary cell of the $fcc$
lattice in which there are two lattice sites in the cases of $Ni$ and $Pd$
investigated.

\subsection{Approximations, identities and relations in calculation of cross
section}

Now we deal with the energy denominator $\left( \Delta E_{\mu i}\right) $ in 
$\left( \ref{Wfi22}\right) $ and $\left( \ref{sigma}\right) $ $\left[ \text{%
see }\left( \ref{DeltaEmui}\right) -\left( \ref{E1f}\right) \right] $. The
shielding parameter $\lambda $ is determined by the innermost electronic
shell of the atom $_{Z}^{A_{1}}X$ and it can be determined as%
\begin{equation}
\lambda =\frac{Z}{a_{B}},  \label{lambda}
\end{equation}%
where $a_{B}=0.53\times 10^{-8}$ $cm$ is the Bohr-radius. The integrals in $%
\left( \ref{Wfi22}\right) $ and $\left( \ref{sigma}\right) $ have
accountable contributions if%
\begin{equation}
\ \left\vert \mathbf{k}_{1}+\mathbf{k}_{2}\right\vert \lesssim \lambda
\label{condlambda}
\end{equation}%
and then $E_{fe}\lesssim \hbar ^{2}\lambda ^{2}/\left( 2m_{e}\right) =\frac{1%
}{2}\alpha _{f}^{2}m_{e}c^{2}Z^{2}$ which can be neglected in $\Delta E_{\mu
i}$ and in the energy Dirac-delta. Thus%
\begin{equation}
\Delta E_{\mu i}=\frac{\hbar ^{2}\mathbf{k}_{1}^{2}}{2m_{1}}+\frac{\hbar ^{2}%
\mathbf{k}_{2}^{2}}{2m_{n}}-\Delta _{-}+\Delta _{n}  \label{DeltaEmui2}
\end{equation}%
and in the Dirac-delta%
\begin{equation}
E_{f}=\frac{\hbar ^{2}\mathbf{k}_{1}^{2}}{2m_{1}}+\frac{\hbar ^{2}\mathbf{k}%
_{2}^{2}}{2m_{2}}.  \label{Ef2}
\end{equation}%
In this case $\mathbf{k}_{1}=-\mathbf{k}_{2}+\delta \mathbf{k}$ with $%
\left\vert \delta \mathbf{k}\right\vert =\delta k\sim \lambda $. Using 
\begin{equation}
k_{1}\simeq k_{2}\simeq k_{0}=\sqrt{2\mu _{12}\Delta }/\hbar  \label{k0}
\end{equation}%
(see below) with $\mu _{12}c^{2}=A_{12}m_{0}c^{2}$, where $A_{12}=\left(
A_{1}-1\right) \left( A_{2}+1\right) /\left( A_{1}+A_{2}\right) $ is the
reduced nucleon number, one can conclude that the $\mathbf{k}_{2}=-\mathbf{k}%
_{1}$ relation fails with a very small error in the cases of events which
fulfill condition $\left( \ref{condlambda}\right) $ since $k_{1}/k_{0}\simeq
1$, $k_{2}/k_{0}\simeq 1$,$\ \delta k/k_{0}\sim \lambda /k_{0}$ and $\lambda
/k_{0}=\alpha _{f}Zm_{e}c^{2}/\sqrt{2\mu _{12}c^{2}\Delta }\ll 1$.
Consequently, the quantity $E_{f}$ in the argument of the energy Dirac-delta
can be written approximately as 
\begin{equation}
E_{f}=\left( \frac{\hbar ^{2}}{2m_{1}}+\frac{\hbar ^{2}}{2m_{2}}\right) 
\mathbf{k}_{2}^{2}=\frac{\hbar ^{2}c^{2}\mathbf{k}_{2}^{2}}{2A_{12}m_{0}c^{2}%
}.  \label{Ef22}
\end{equation}%
Furthermore taking $A_{1}/\left( A_{1}+1\right) \simeq 1$ 
\begin{equation}
\Delta E_{\mu i}=\frac{\hbar ^{2}c^{2}\mathbf{k}_{2}^{2}}{2m_{0}c^{2}}%
-\Delta _{-}+\Delta _{n}.  \label{DeltaEmui3}
\end{equation}

We introduce the $\mathbf{Q}=\hbar c\mathbf{k}_{2}/\Delta $, $\mathbf{P}%
=\hbar c\left( \delta \mathbf{k}\right) /\Delta $, $\varepsilon
_{f}=E_{f}/\Delta =\left[ \mathbf{Q}^{2}/\left( 2A_{12}m_{0}c^{2}\right) %
\right] \Delta $ and $L=\hbar c\lambda /\Delta $ dimensionless quantities.
The energy Dirac-delta modifies as $\delta (E_{f}-\Delta )=\delta \left[
\varepsilon _{f}\left( \mathbf{Q}\right) -1\right] /\Delta $. \ The relation 
$\left( \ref{lambda}\right) $ yields $L=\hbar cZ/\left( a_{B}\Delta \right)
=Z\alpha _{f}m_{e}c^{2}/\Delta $ and $Z\alpha _{f}m_{e}c^{2}/\Delta \lesssim
1$. Now we change $d^{3}k_{1}d^{3}k_{2}$ to $\left( \frac{\Delta }{\hbar c}%
\right) ^{6}d^{3}Qd^{3}P$ in the integration in $\left( \ref{sigma}\right) $%
, use the $\delta \left[ g\left( Q\right) \right] =\delta \left(
Q-Q_{0}\right) /g^{\prime }\left( Q_{0}\right) $ identity, where $Q_{0}$ is
the root of the equation $g\left( Q\right) =0$ ($k_{0}=Q_{0}\Delta /\left(
\hbar c\right) $, see $\left( \ref{k0}\right) $), estimate the integral with
respect to the components of $\mathbf{P}$ by 
\begin{equation}
\int_{0}^{\infty }\frac{4\pi P^{2}dP}{\left( P^{2}+L^{2}\right) ^{2}}=\frac{%
\pi ^{2}}{L}  \label{IntP}
\end{equation}%
and apply $v_{c}=d^{3}/4$ (the volume of unit cell of $fcc$ lattice for $Ni$
and $Pd$ of lattice parameter $d$).

\subsection{$\left\langle \left\vert F_{1}\left( \mathbf{k}_{0}\right)
\right\vert ^{2}\right\rangle _{Sh}$ and $\sum_{l_{2}=-m_{2}}^{l_{2}=m_{2}}%
\left\vert F_{2}\left( \mathbf{k}_{0}\right) \right\vert _{Sh}^{2}$\ in
single particle shell-model and without LWA}

Now we calculate the quantities $\left\langle \left\vert F_{1}\left( \mathbf{%
k}_{0}\right) \right\vert ^{2}\right\rangle _{Sh}$~and $%
\sum_{l_{2}=-m_{2}}^{l_{2}=m_{2}}\left\vert F_{2}\left( \mathbf{k}%
_{0}\right) \right\vert _{Sh}^{2}$ in the single particle shell model with
isotropic harmonic oscillator potential and without the long wavelength
approximation (see definitions: $\left( \ref{F1kalk2}\right) $, $\left( \ref%
{F1av}\right) $ and $\left( \ref{F2k2}\right) $). Taking into account the
spin-orbit coupling in the level scheme the emerging neutron states are $0l$
and $1l$ shell model states in the cases of $Ni$ and $Pd$ to be discussed
numerically \cite{Pal}. So the initial and final neutron states $\left( \Phi
_{i1},\Phi _{f2}\right) $ have the form%
\begin{equation}
\Phi _{Sh}\left( \mathbf{r}_{nj}\right) =\frac{R_{n_{j}l_{j}}}{r_{nj}}%
Y_{l_{j}m_{j}}\left( \Omega _{j}\right)  \label{Fishell}
\end{equation}%
where $n_{j}=0,1$ in the cases of $0l$ and $1l$ investigated, respectively,
and 
\begin{equation}
R_{0l_{j}}=b_{j}^{-1/2}\left( \frac{2}{\Gamma (l_{j}+3/2)}\right)
^{1/2}\varrho _{j}^{l_{j}+1}\exp \left( -\frac{1}{2}\varrho _{j}^{2}\right) ,
\label{R0l}
\end{equation}%
\begin{eqnarray}
R_{1l_{j}} &=&b_{j}^{-1/2}\left( \frac{2l_{j}+3}{\Gamma (l_{j}+3/2)}\right)
^{1/2}\varrho _{j}^{l_{j}+1}\times  \label{R1l} \\
&&\times \left( 1-\frac{2}{2l_{j}+3}\varrho _{j}^{2}\right) \exp \left( -%
\frac{1}{2}\varrho _{j}^{2}\right)  \notag
\end{eqnarray}%
with $\varrho _{j}=r_{nj}/b_{j}$ where $b_{j}=\sqrt{\hbar /\left(
m_{0}\omega _{j}\right) }$ \cite{Pal}. Here $\omega _{j}$ is the angular
frequency of the oscillator that is determined by $\hbar \omega
_{1}=40A_{1}^{-1/3}$ $MeV$ and\ $\hbar \omega _{2}=40\left( A_{2}+1\right)
^{-1/3}$ $MeV$ \cite{Bohr}. (The subscript $Sh$ refers to the shell model.)
With the aid of these wave functions and for $n_{1}=0,1$ 
\begin{equation}
\left\langle \left\vert F_{1}\left( \mathbf{k}_{0}\right) \right\vert
^{2}\right\rangle _{Sh}=b_{1}^{3}\frac{2^{l_{1}+2}}{\sqrt{\pi }\left(
2l_{1}+1\right) !!}4\pi I_{1,n_{1}}^{2}  \label{F1K2Sh}
\end{equation}%
with 
\begin{equation}
I_{1,0}=\int_{0}^{\infty }\varrho ^{l_{1}+2}j_{l_{1}}(k_{0}b_{1}\frac{A_{1}}{%
A_{1}-1}\varrho )e^{-\frac{1}{2}\varrho ^{2}}d\varrho \text{ }  \label{I10}
\end{equation}%
and 
\begin{eqnarray}
I_{1,1} &=&\left( l_{1}+\frac{3}{2}\right) \int_{0}^{\infty }\varrho
^{l_{1}+2}\left( 1-\frac{2}{2l_{1}+3}\varrho ^{2}\right) \times  \label{I11}
\\
&&\times j_{l_{1}}(k_{0}b_{1}\frac{A_{1}}{A_{1}-1}\varrho )e^{-\frac{1}{2}%
\varrho ^{2}}d\varrho .\text{ }  \notag
\end{eqnarray}%
Here $j_{l_{1}}(x)=\sqrt{\frac{\pi }{2x}}J_{l_{1}+1/2}(x)$ denotes spherical
Bessel function with $J_{l_{1}+1/2}(x)$ the Bessel function of first kind.

Similarly%
\begin{eqnarray}
\sum_{l_{2}=-m_{2}}^{l_{2}=m_{2}}\left\vert F_{2}\left( \mathbf{k}%
_{0}\right) \right\vert _{Sh}^{2} &=&b_{2}f^{2}\frac{2^{l_{2}+2}\left(
2l_{2}+1\right) }{\sqrt{\pi }\left( 2l_{2}+1\right) !!}\times  \label{F2k2Sh}
\\
&&\times 4\pi \left( \frac{A_{2}}{A_{2}+1}\right) ^{2}I_{2,n_{2}}^{2}  \notag
\end{eqnarray}%
with%
\begin{equation}
I_{2,0}=\int_{0}^{\infty }\varrho ^{l_{2}+1}j_{l_{2}}(k_{0}b_{2}\varrho )e^{-%
\frac{1}{2}\varrho ^{2}-\frac{A_{2}+1}{A_{2}}\frac{b_{2}}{r_{0}}\varrho
}d\varrho \text{ }  \label{I20}
\end{equation}%
and%
\begin{eqnarray}
I_{2,1} &=&\left( l_{2}+\frac{3}{2}\right) \int_{0}^{\infty }\varrho
^{l_{2}+1}\left( 1-\frac{2}{2l_{2}+3}\varrho ^{2}\right) \times  \label{I21}
\\
&&\times j_{l_{2}}(k_{0}b_{2}\varrho )e^{-\frac{1}{2}\varrho ^{2}-\frac{%
A_{2}+1}{A_{2}}\frac{b_{2}}{r_{0}}\varrho }d\varrho .  \notag
\end{eqnarray}%
Substituting the results of $\left( \ref{F1K2Sh}\right) $, $\left( \ref%
{F2k2Sh}\right) $ and $\left( \ref{F1K2av}\right) $ into $\left( \ref{etha}%
\right) $ one gets%
\begin{eqnarray}
\eta _{l_{1},n_{1},l_{2},n_{2}}\left( A_{1},A_{2}\right) &=&\frac{%
2^{l_{1}+l_{2}+4}}{\pi \left( 2l_{1}+1\right) !!\left( 2l_{2}+1\right) !!}%
\times  \label{etha2} \\
&&\times \frac{b_{1}^{3}b_{2}}{r_{0}^{4}}\left( \frac{A_{2}}{A_{2}+1}\right)
^{2}I_{1,n_{1}}^{2}I_{2,n_{2}}^{2}.  \notag
\end{eqnarray}

\end{document}